\documentclass[%
preprint,
 amsmath,amssymb,
 aps,
]{revtex4-1}

\usepackage{graphicx}
\usepackage{dcolumn}
\usepackage{bm}
\usepackage{subfig}%
\usepackage{url}%
\usepackage{color,soul}%

\begin{document}

\preprint{APS/123-QED}

\title{Manuscript Title:\\Tailorable Twisting of Biomimetic Scale-Covered Substrate}

\author{Hossein Ebrahimi, Hessein Ali, Ryan Alexander Horton, Jonathan Galvez, Ali Gordon and Ranajay Ghosh}
 \email{ranajay.ghosh@ucf.edu}
 \homepage{http://mae.ucf.edu/cosmos/}




\affiliation{%
 Department of Mechanical and Aerospace Engineering, University of Central Florida, Orlando FL, 32816
}%

\date{\today}

\begin{abstract}
In this letter, we investigate the geometrically tailorable elasticity in the twisting behavior of biomimetic scale-covered slender soft substrate. Motivated by our qualitative experiments showing a significant torsional rigidity increase, we develop an analytical model and carry out extensive finite element (FE) simulations to validate our model. We discover a regime differentiated and reversible mechanical response straddling linear, nonlinear, and rigid behavior. The response is highly tailorable through the geometric arrangement and orientation of the scales. The work outlines analytical relationships between geometry, deformation modes and kinematics, which can be used for designing biomimetic scale-covered metamaterials.
\end{abstract}

\maketitle

Scales have been a recurring dermal feature in the evolutionary history of complex vertebrae ~\cite{c01,c02,c03,c04,c05,c06,c07}. One of the reasons for their success is their tremendous multifunctional benefits, including hydrodynamic, chemical, and optical advantages ~\cite{c08,c09,c10,c11}. From a mechanical standpoint, scales traditionally provide protection against foreign objects and organism attacks  ~\cite{c12,c13}. This evolutionary requirement has made them structurally hybrid  ~\cite{c14,c15,c16}, hierarchical ~\cite{c17,c18,c19}, and composite in nature ~\cite{c20,c21,c22} capable of engaging multiple length scales ~\cite{c23,c24,c25,c26,c270,c27,c28,c29}. This feature has been an inspiration for recent work on using these principles for armor designs ~\cite{c06,c12,c26,c30}. However, in addition to localized loads common in protective applications, another structural feature is of immense importance. That is the role of scale engagements in modifying the global deformation behavior of the underlying structure. This feature deepens the role of scales in aiding both locomotion  ~\cite{c31} and swimming  ~\cite{c32,c33}. The mechanics underscoring this behavior have been an area of intense scrutiny since the last few years. One-dimensional substrates with stiff scales revealed strain stiffening due to sliding, scale deformation as well as friction in the bending mode of deformation ~\cite{c34}. Later simplification of the structure revealed the distinct nonlinear regimes of elasticity even without scale deformation or friction ~\cite{c35}. Nonlinearity due to frictional effects were further isolated and their effect on locking and dissipation quantified ~\cite{c36}. Further studies revealed the limits of theoretical assumptions underpinning the models and their effect on predicted relationships ~\cite{c37,c38}. Extending the dimensionality of the problem, two-dimensional substrates were also investigated ~\cite{c39,c40,c41,c42}. These showed several similarities with their one-dimensional counterparts in bending. However, the mechanics of twisting, a critical and fundamental mode of global deformation has not yet been discussed in detail. Twisting mode of deformation is perceptible for some robotic applications ~\cite{c43,c44,c45,c46,c47,c48,c49} and can also arise due to boundary defects and bending-twisting coupling in structures ~\cite{c50,c51}. Furthermore, this mode is an important first step towards more complex combined deformation modes and two-dimensional metamaterials of this type. In this letter, we study the response of stiff scale-covered slender biomimetic substrates under torsional loads and outline the gamut of geometrical tailorable of twisting elasticity. 

\begin{figure}[htbp]
\centering
\subfloat[]{%
\includegraphics[scale =0.117]{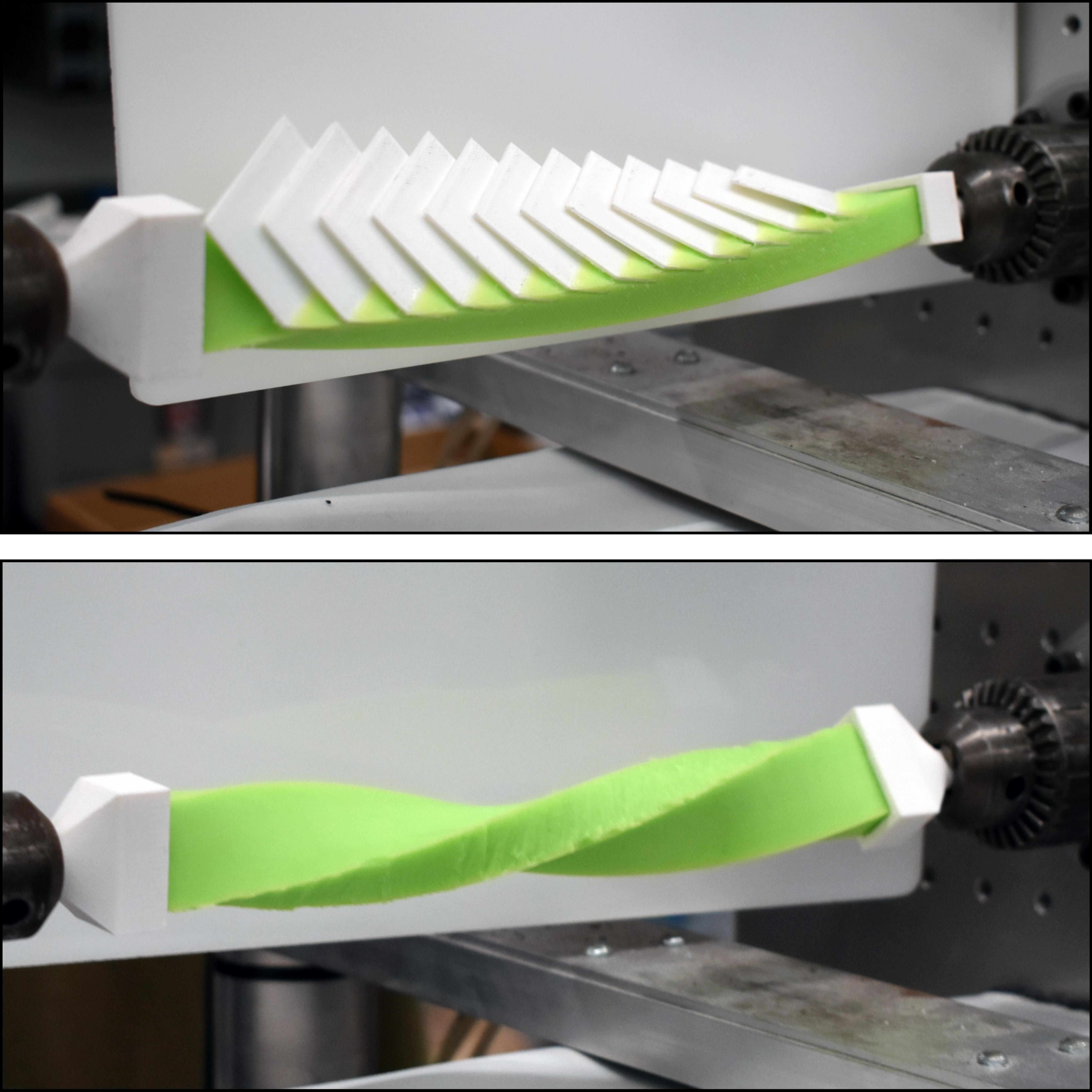}
\label{fig1A}}
\quad
\subfloat[]{%
\includegraphics[scale=0.31]{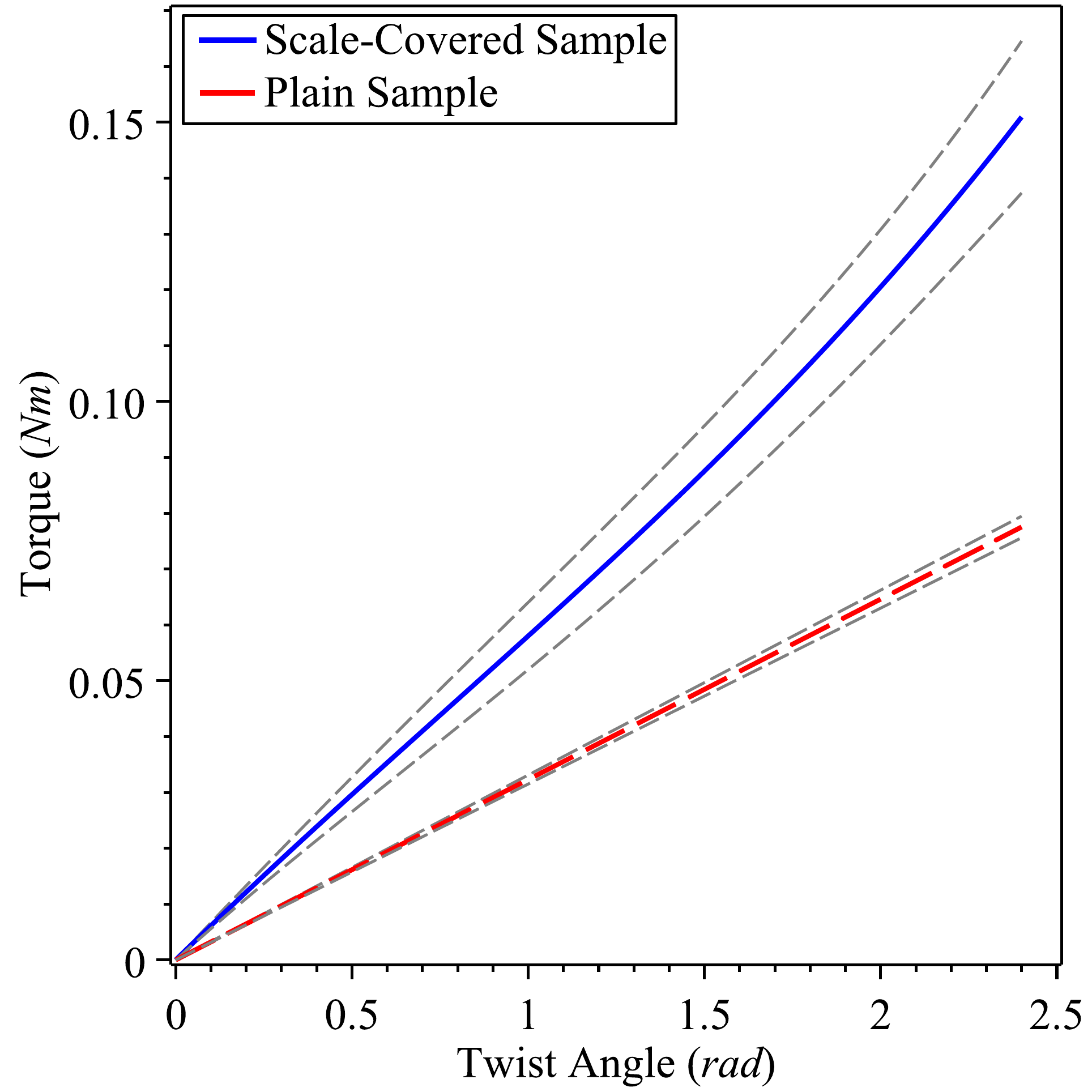}
\label{fig1B}}
\caption{(a) Qualitative twisting experiments on the scale-covered and plain beams. (b) Exhibiting nonlinear torque-twist angle plots with uncertainties shown in dashed lines. Each substrate's dimensions were $200 \times 25 \times 12.5$ $mm$, scales dimensions were $36 \times 36 \times 1$ $mm$ with a spacing of $d=14$ $mm$ and inclusion length of $L=10$ $mm$. Initial scale inclination angle and alpha are $\theta_0=10^\circ$ and $\alpha =45^\circ$, respectively. These geometrical parameters are shown in Fig.~\ref{fig2}.}
\label{fig1} 
\end{figure}

Twisting deformation of a uniform prismatic beam covered with scales and a plain sample is shown in Fig.~\ref{fig1A}. We also carry on qualitative and motivational experiments on some real samples. For these samples, the scales were 3D printed and made of Poly Lactic Acid (PLA) thermoplastic. The substrate was made from a silicone elastomer known as Vinylpolysiloxane (VPS) by casting into a 3D printed mold. The mold was designed to leave grooves for scale insertions at the next stage. Then the scales were then embedded and adhered to these prefabricated grooves, a silicone glue (Permatex Corporate). Young's modulus of PLA and VPS tested under tensile test by MTS Insight\textregistered~ (Electromechanical – $50$ $kN$ Standard Length) were found to be $2.86$ $GPa$ and $1.5$ $MPa$, respectively. We subjected the fabricated samples to twisting experiments in clockwise direction, contrasting their response using an MTS Bionix EM\textregistered~ (Electromechanical Torsion – $45$ $Nm$) with similar loading and boundary conditions. Note that the engagement happens only in the clockwise twist direction of the substrate. The twist rate was $0.085$ $RPM$ and the experiments have been done up to $2.4$ $rad$. The significant gains in torsional stiffness in the scale-covered samples were apparent when compared to a plain counterpart with same materials and geometry as shown in Fig.~\ref{fig1B}. The uncertainties from different tests are shown as dashed lines up and down of each curves.

\begin{figure}[htbp]
\centering
\subfloat[]{%
\includegraphics[scale = 0.34]{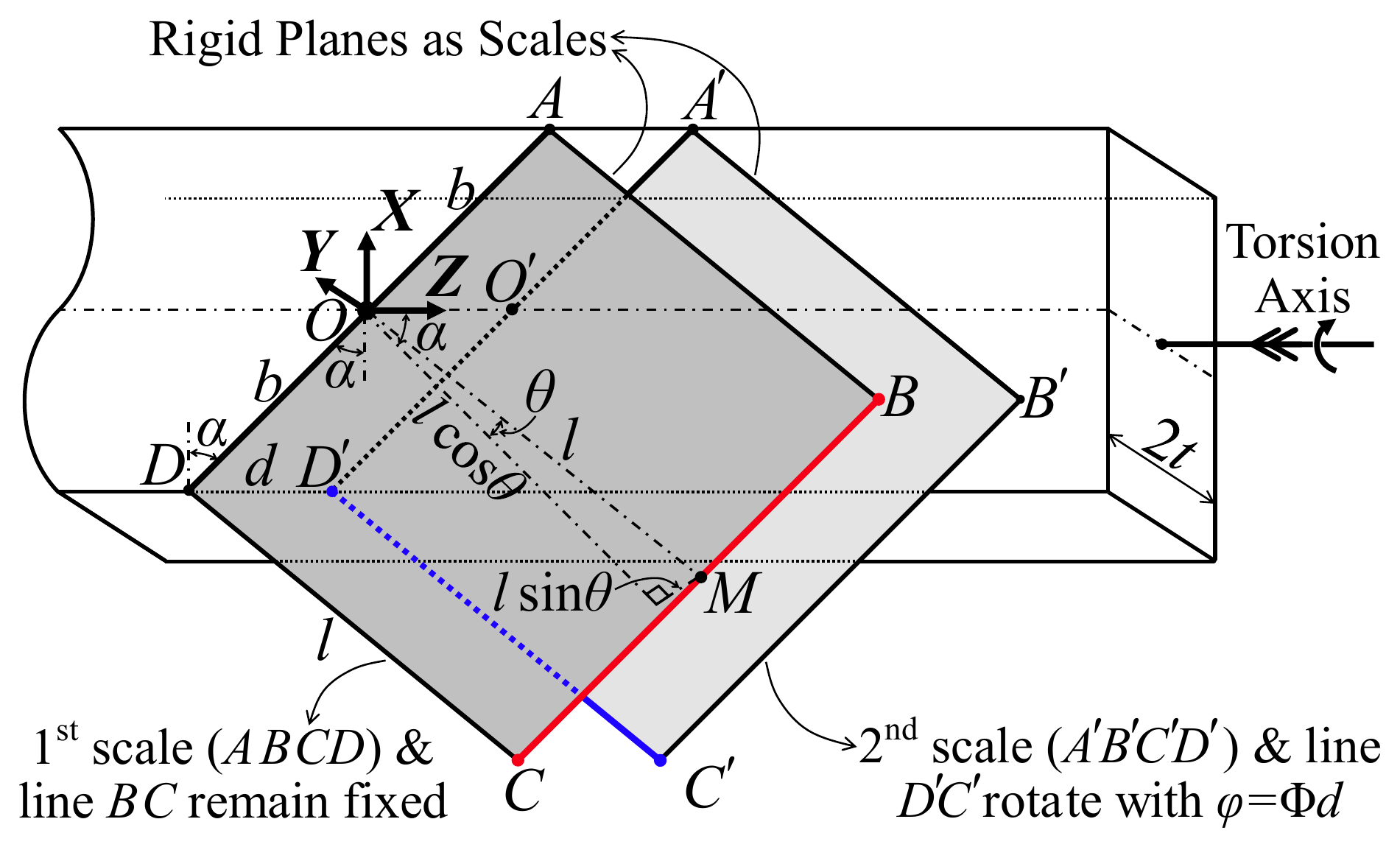}
\label{fig2A}}
\quad
\subfloat[]{%
\includegraphics[scale=0.34]{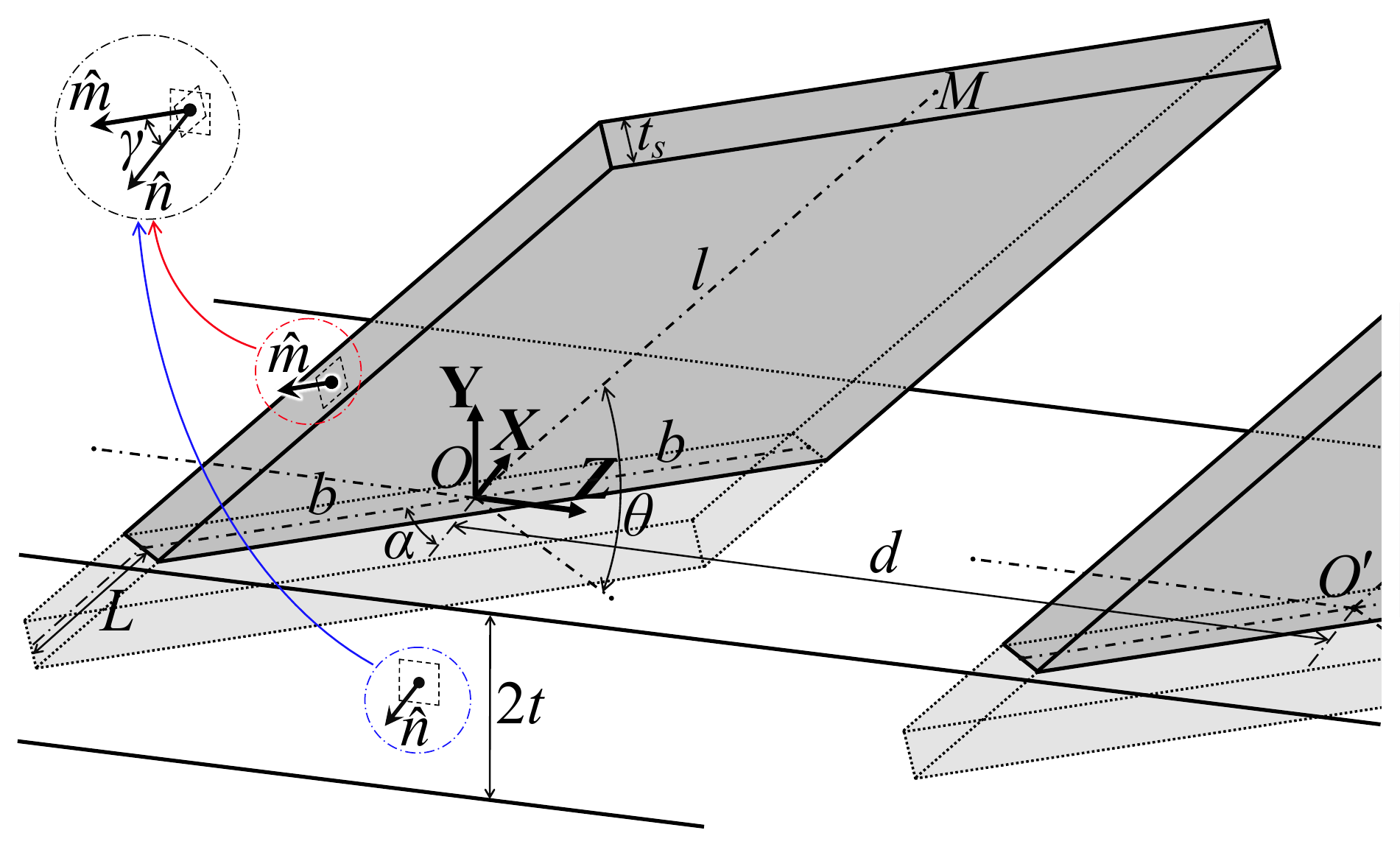}
\label{fig2B}}
\caption{(a) Perspective top view of geometrical configuration between two consecutive scales in a general form under torsional load. (b) Dimetric view of (a), showing the three orientational angles $\theta$, $\alpha$, and $\gamma$. $\hat{m}$ and $\hat{n}$ are the normal unit vectors of the scale's side surface and substrate's side surface, respectively.}
\label{fig2} 
\end{figure}

Using this motivating experiment, we investigate this twisting behavior by first developing a model for this system. Due to the high contrast in the elastic modulus between substrate and scale stiffness, we use a rigid scale assumption. We first simplify the geometry of this system, Fig.~\ref{fig2A}. Each scale is considered a rectangular plate with thickness $t_s$, width $2b$ and total length $l_s=L+l$, where $L$ is the length of the substrate embedded section and $l$ is the length of the exposed section. The patterned row of scales, spaced by $d$ and embedded on a rectangular prismatic substrate, is quantified in a general orientation defined with three angles $\theta$, $\alpha$, and $\gamma$, Fig.~\ref{fig2B}. Here, $\theta$ is the dihedral angle between the top surfaces of the substrate and the scale, $\alpha$ is the angle between the substrate's rectangular cross section and the edge of the scale’s width, and $\gamma$ is the dihedral angle between the side surfaces of the substrate and the scale (the angle between normal unit vectors $\hat{m}$ and $\hat{n}$ of the side surfaces shown in Fig.~\ref{fig2B}). We assume that the length of the beam and number of scales are large enough to satisfy the periodicity in the scale engagement under pure torsion with negligible edge effects. This allows the isolation of a representative volume element (RVE) formed one scale and the underlying substrate (scale's thickness is neglected) at distance $d$ from adjacent scales embedded on the top surface of the beam, Fig.~\ref{fig2A}.

To develop the kinematics of the scale at the RVE level, consider the RVE scale (1st scale in Fig.~\ref{fig2A}) fixed with respect to the one immediately preceding it (2nd scale in Fig.~\ref{fig2A}). The second scale rotates about the torsion axis by twist angle of $\varphi$ caused by the twisting of the underlying slender substrate, Fig.~\ref{fig2A}. This leads to an RVE (local) twist rate $\Phi={\varphi \over d}$. From Fig.~\ref{fig2A}, if the second scale twists around the torsion axis, then engagement would only happen when one edge of rotated scale, $D'C'$, contacts at a point with the subsequent edge $BC$ of the fixed scale. The continual twisting of the underlying beam progresses the contact between two consecutive scales, increasing the scale's inclination angle from an initial angle $\theta_0$ to the current angle $\theta$. This contact imposes kinematic constraints on scale sliding, leading to the following nonlinear relationship between the substrate's twist angle $\varphi$ and scale's inclination angle $\theta$ (See Supplemental Material for derivation ~\cite{c52}):

\begin{multline}
  \left( {\cos \varphi  - 1} \right)\left[ {\beta \sin 2\alpha \sin \theta  + \eta {{\cos }^2}\alpha \sin 2\theta  + 2\lambda \cos 2\alpha \cos \theta } \right]   \\
    + 2\sin \alpha \sin \varphi \left( {\eta  + \lambda \sin \theta } \right) + 2\cos \alpha \sin \varphi \cos \theta \left( {\beta  - \sin \alpha } \right) - 2\cos \alpha \cos \varphi \sin \theta  = 0. 
\label{eq1}
\end{multline}

Where $\eta=l/d$, $\beta=b/d$, and $\lambda=t/d$ are the overlap ratio, dimensionless scale width, and dimensionless substrate thickness, respectively. In the small twisting regime $(\theta \ll 1, \varphi \ll 1)$, the implicit constraint equation simplifies to the explicit $\theta=\varphi(\eta \tan \alpha +\beta- \sin \alpha)$. Furthermore for $\alpha, \beta \ll 1$ (thin substrate with grazing scales), we get $\theta \sim \varphi(\eta \alpha +\beta- \alpha)$. For higher $\eta$ range, i.e. $\eta \gg 1$, then for fixed $\alpha$ and $\beta$ the first term will dominate and we will get $\theta \sim \eta \varphi$. This is similar to the scaling law obtained from bending deformation ~\cite{c35} and underscores the universal importance of scale overlap in amplifying global to local deformation. However, unlike bending, we also find an additional amplification factor $\theta \sim \beta \varphi$ underscoring the more general nature of this system.

\begin{figure}[h]
\centering
\subfloat[]{%
\includegraphics[scale = 0.34]{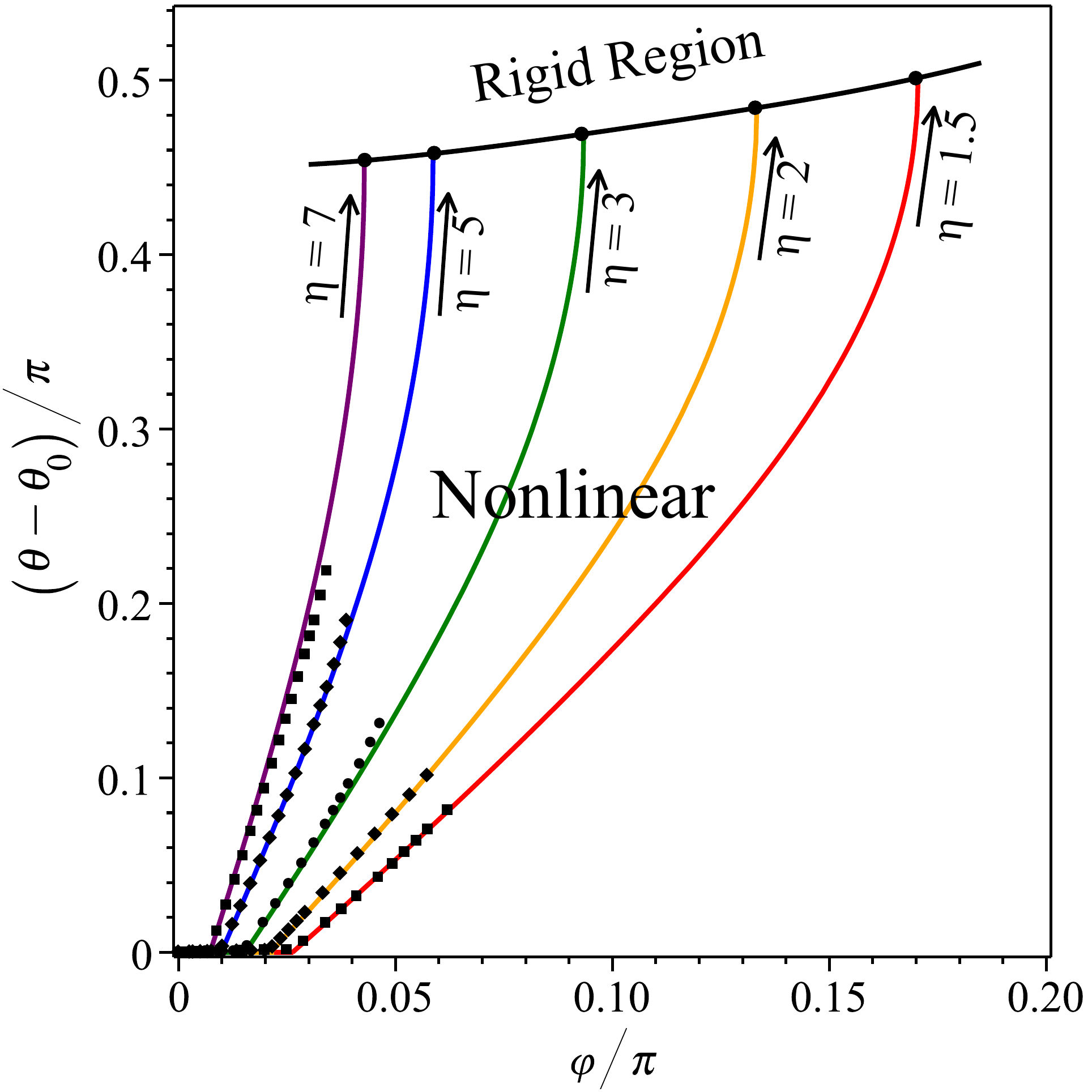}
\label{fig3A}}
\quad
\subfloat[]{%
\includegraphics[scale=0.34]{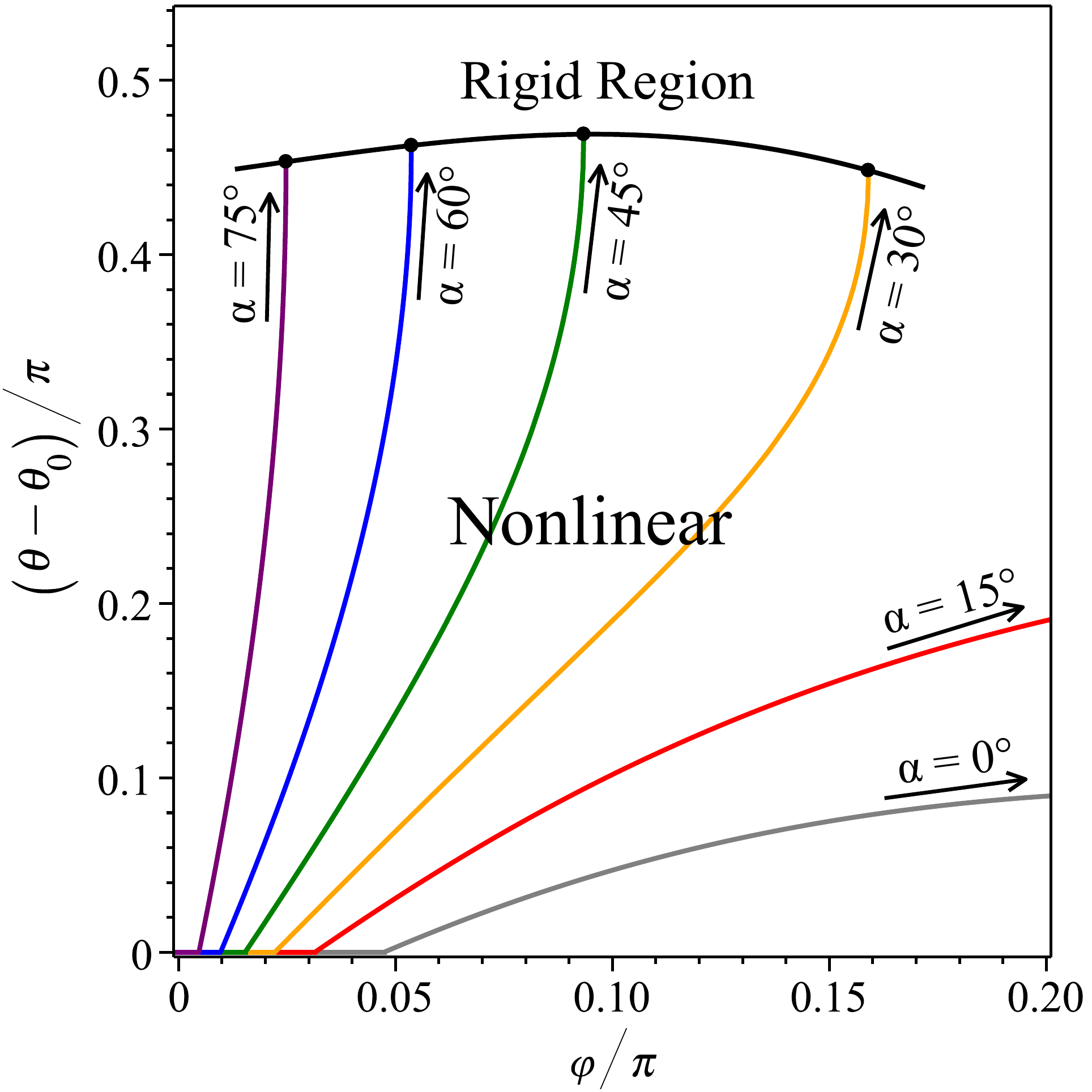}
\label{fig3B}}
\caption{(a) Phase map of the biomimetic system under torsional loading includes three distinct regions of operation: linear, nonlinear, and rigid, for different $\eta$ with the given values of $\alpha=45^\circ$, $\theta_0=10^\circ$, $\beta=1.25$, and $\lambda=0.45$. Black dotted lines represent FE results. (b) Phase map of the system for different $\alpha$ with the given values of $\eta=3$, $\theta_0=10^\circ$, $\beta=1.25$, and $\lambda=0.45$.}
\label{fig3} 
\end{figure}

The governing nonlinear Eq.~\ref{eq1} establishes a phase map spanned by $(\theta-\theta_0)/\pi$ and $\varphi/\pi$, which is shown in Fig.~\ref{fig3A} for different $\eta$ along with FE simulation results ~\cite{c52}. Here, $\theta_0=10^\circ$, $\alpha=45^\circ$, $\beta=1.25$, and $\lambda=0.45$. This phase diagram maps out three kinematic regimes of operations for the structure under twisting, which includes linear, nonlinear and rigid regions of operation. The linear region is a direct result of the non-engagement of scales. However, as soon as the scales begin engaging, a distinctly nonlinear regime emerges, tuned by $\eta$. The angle of engagement $\varphi_e$ decreases with increasing the overlap ratio. For relatively smaller deformation, an explicit relationship emerges between $\varphi_e$ and other kinematic parameters, $\varphi_e={\theta_0 \over {\eta \tan \alpha+ \beta- \sin \alpha}}$. The stiffening increases with scale sliding, ultimately leading to a point where no more sliding is possible without significantly deforming the scales themselves. This is the third regime of deformation, called ''locking''. At this point, the system begins to behave almost as a rigid body, completing the regime traversal. We find this rigidity envelope mathematically by satisfying $\partial \varphi /\partial \theta=0$, which forms the defining envelope of operation. Locking signals a sharp rise in contact forces, which violates the scale rigidity condition near the envelope due to local scale deformation. Around that phase boundary, the stiffness of the whole system transitions towards the stiffness of the scales, which are significantly stiffer than the substrate. This is consistent with previously published work on this topic ~\cite{c35,c41}.

Also, if $\eta$ is sufficiently small, no engagement is possible and this is the geometrical limit of engagement. This computes to $\eta_c={{1-2\beta \sin \alpha}\over{\cos \alpha \cos \theta_0}}$ and is physically meaningful if $\eta_c>0$ ~\cite{c52}. The agreement of the analytical relationship with FE simulations in Fig.~\ref{fig3A} shows minimal effect of substrate warping on the kinematics.

We explore this geometric tailorability of elasticity in greater detail using another phase map, parametrized by $\alpha$, Fig.~\ref{fig3B} with $\eta=3$, $\theta_0=10^\circ$, $\beta =1.25$, and $\lambda=0.45$. This phase map shows that increasing $\alpha$ not only leads to a quicker engagement but also steeper nonlinearity. This also shows that there exists a critical $\alpha_c$, below which no locking would be possible for a given set of geometrical parameters. Also note that, although increasing $\eta$ always leads to decreasing locking twist angle, this trend does not hold for $\alpha$. 

\begin{figure}[htbp]
\centering
\subfloat[]{%
\includegraphics[scale =0.3]{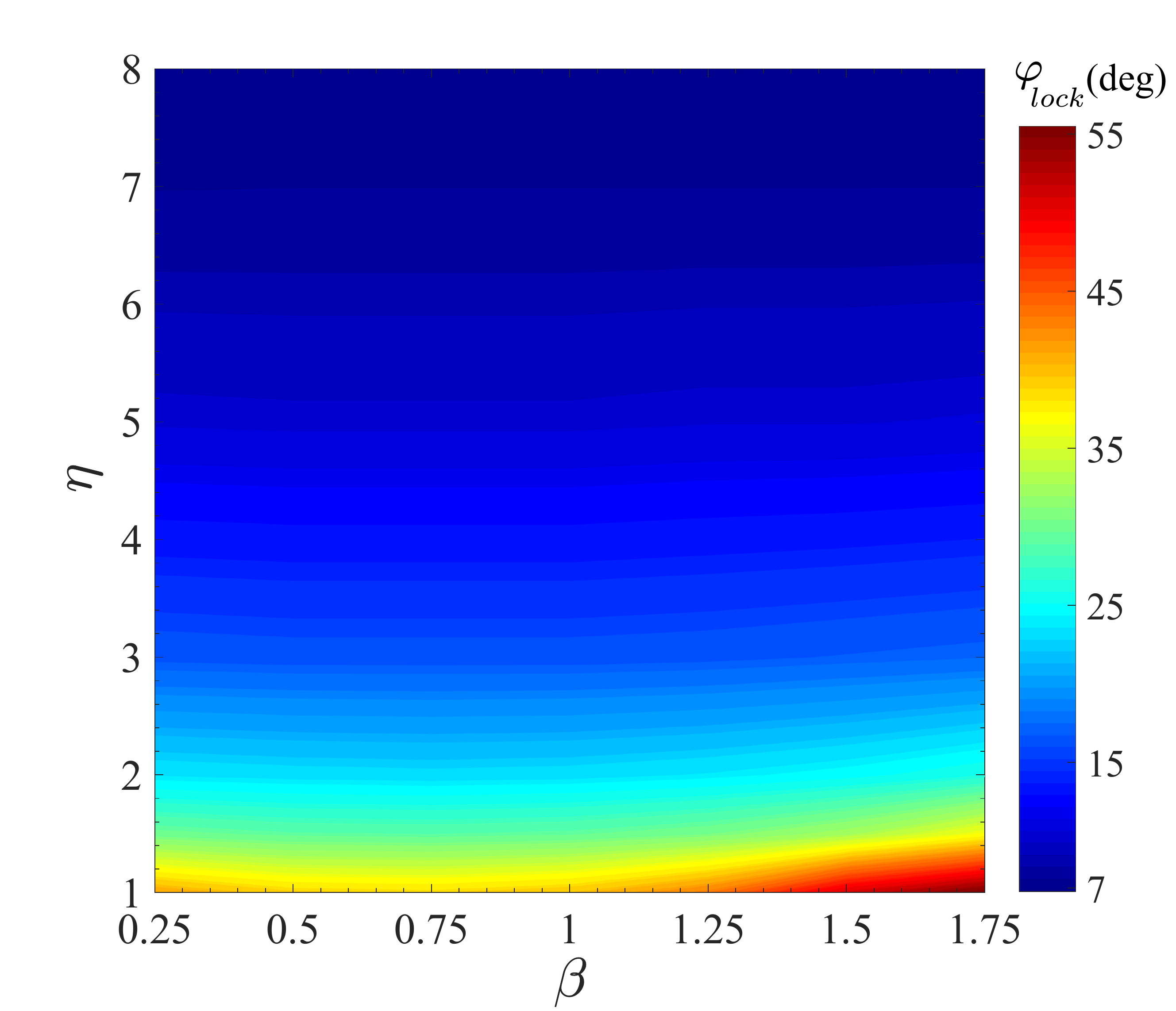}
\label{fig4A}}
\quad
\subfloat[]{%
\includegraphics[scale=0.3]{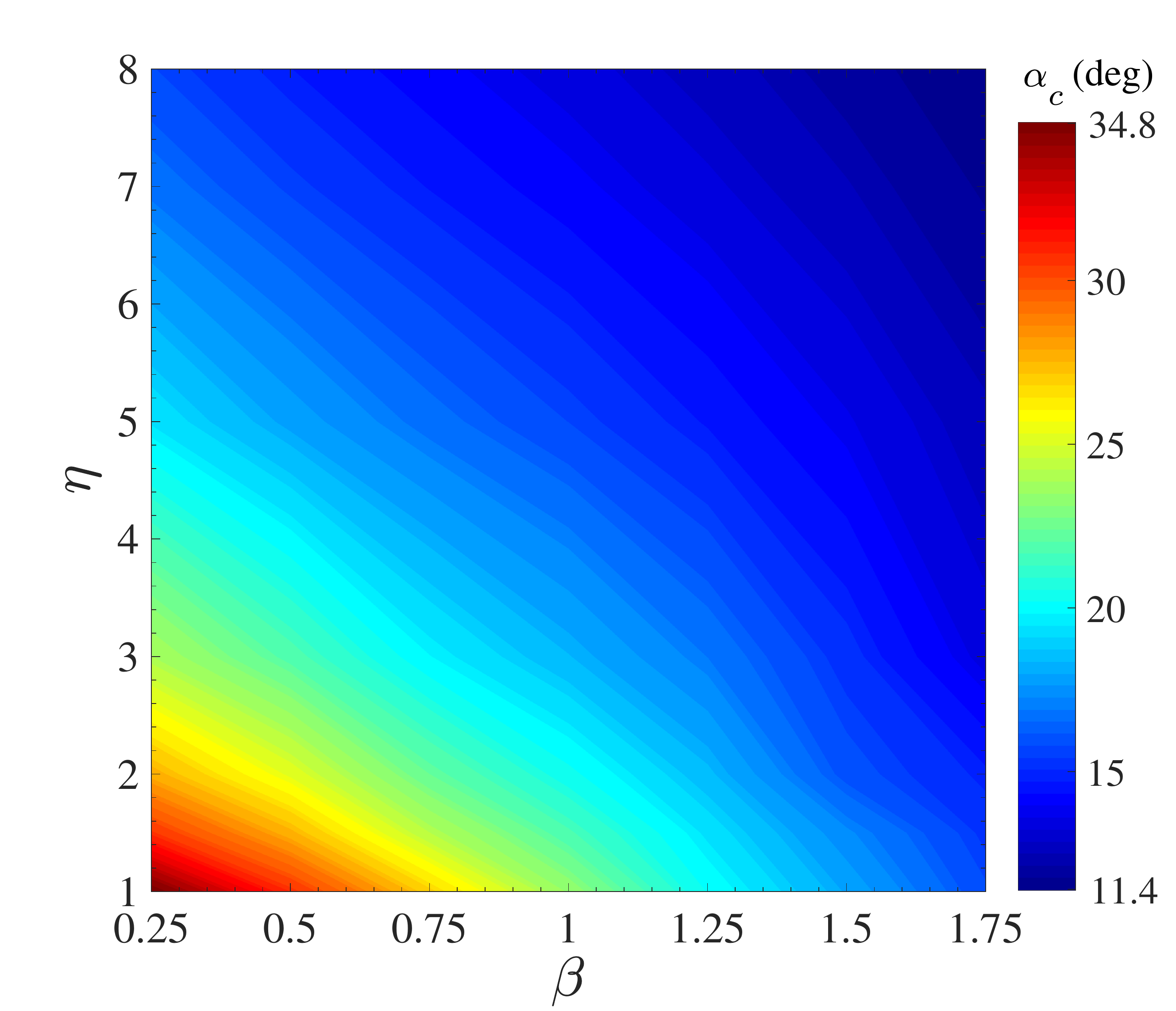}
\label{fig4B}}
\caption{(a) Phase plot of locking $\varphi$ ($\varphi_{lock}$), spanned by $\eta$ and $\beta$ with the given values of $\alpha=45^\circ$, $\theta_0=10^\circ$ and $\lambda=0.45$. (b) Phase plot of critical $\alpha$ ($\alpha_c$), spanned by $\eta$ and $\beta$ with the given values of $\theta_0=10^\circ$ and $\lambda=0.45$.}
\label{fig4} 
\end{figure}

These effects are summarized using two other phase diagrams, both spanned by $\eta$ and $\beta$ in Fig.~\ref{fig4A} and Fig.~\ref{fig4B}. Fig.~\ref{fig4A} indicates that locking angles decrease for higher $\eta$. Further, higher $\eta$ depresses the sensitivity of locking angle to $\beta$. However, the locking angle sharply increases with $\beta$ for low enough $\eta$. Therefore, only in the lower overlap ratios, $\beta$ becomes an important tuning parameter of the locking behavior. Interestingly, this phase plot shows that although higher $\eta$ always decreases the envelope of operation, the influence of $\beta$ is strongly dependent on $\eta$. In Fig.~\ref{fig4B}, which tracks the critical angle $\alpha_c$ below which locking would not take place, similar tuning behavior of $\beta$ is apparent. However, in this case, as $\beta$ increases the locking possibility improves. Note that, in these phase plots $\eta_c<1$. 

These kinematic nonlinearities ensure that mechanical response would also be nonlinear even when the materials themselves are in the linear elastic regime. We consider the twisting of the biomimetic scale-covered substrate as a combination of plain beam twisting and scales rotation. The scale rotation in 3D space can be defined by change in angles $\theta$, $\alpha$, and $\gamma$, introduced earlier, Fig.~\ref{fig2B}. As the scales engage and begin rotating, the elastic substrate resists scales rotation. The substrate resistance is modeled as linear torsional springs corresponding to the change in each of the angles $\theta$, $\alpha$, and $\gamma$. Thus the energy absorbed due to the 3D rotation of each scale can be described as $U_{scale}={1 \over 2} K_\theta (\theta-\theta_0 )^2+{1 \over 2} K_\alpha (\alpha-\alpha_0 )^2+{1 \over 2} K_\gamma (\gamma -\gamma _0 )^2$, where $K_\theta$, $K_\alpha$, and $K_\gamma$ are the corresponding rotational spring constants. Extensive FE simulations indicate that the contributions from both $K_\alpha$ and $K_\gamma$ terms are negligible ~\cite{c52}. This leads to $U_{scale} \approx {1 \over 2} K_\theta (\theta-\theta_0 )^2$. The most significant variables for determining the scale-substrate joint stiffness $K_\theta$ is the Young's Modulus of substrate $E_B$, the embedded length of scales $L$, width of the scale $2b$, the thickness of the scale $t_s$ as well as $\theta_0$ and $\alpha$. We further assume that $t_s \ll l_s$ and $0 \ll L \ll 2t$. Considering this system as a single scale embedded in a semi-infinite beam in length and thickness, we postulate the following scaling expression:

\begin{equation}
{{{K_\theta }} \over {{E_B}{t_s^2}}} = {C_B}\left( \alpha  \right)b{\left( {{L \over t_s}} \right)^n}f\left( {{\theta _0}} \right),
\label{eq2}
\end{equation}

where $n$ is a dimensionless constant, and $f(\theta_0 )$ and $C_B(\alpha)$ are dimensionless angular functions. We carried out FE simulations on a single scale embedded in a semi-infinite media and varied the relevant geometric variables of Eq.~\ref{eq2} to ascertain the fit of this empirical relationship. We find an excellent fit in the region of $12 < L/t_s < 80$, yielding $n=1.55$, $C_B(\alpha)=3.62$, and $f(\theta_0)\approx 1$, indicating negligible angular dependence ~\cite{c52}. 

In non-circular cross sections, warping leads to an out-of-plane displacement even in small deformation ~\cite{c53}. Although warping’s effect of kinematics was negligible, its effect on mechanics must be accounted. Typically, the effect of warping is addressed using a non-dimensional pre-multiplier $C_w$ in the relationship between torque and twist rate leading to $T=C_w G_B I\Phi$, where $T$ is the RVE (local) torque, $G_B$ is the shear modulus of elasticity, and $I$ is the moment of area of the beam cross section, respectively. $C_w$ can be found from literature for standard cross sections ~\cite{c54}. In addition, the embedding of rigid inclusions leads to an increase in stiffness even before engagement and is modeled using an inclusion correction factor $C_f$, which would depend on the volume fraction, shape and size of the inclusion. This leads to a modified torque-twist relationship $T=C_f C_w G_B I\Phi$.  Motivated by elasticity arguments, we postulate that $C_f=1+C_0 (\alpha)({{\zeta \beta}\over{\lambda}})^m h(\theta_0)$, where $\zeta=L/d$, $h(\theta_0)$ and $C_0 (\alpha)$ are dimensionless angular functions to describe the dependence of $C_f$ to $\theta_0$ and $\alpha$. We ascertained the fit, using FE numerical simulations ~\cite{c52}, yielding $m=1$, $C_0(\alpha)=1.33$, and $h(\theta_0)\approx 1$, indicating negligible angular dependence. With these assumptions, work-energy balance for the unit length of the substrate can be described as:

\begin{equation}
\mathop \smallint \limits_0^{\rm{\Phi }} T\left( {{\rm{\Phi }}'} \right)d{\rm{\Phi }}' = {1 \over 2}{C_f}{C_w}{G_B}I{{\rm{\Phi }}^2} + {1 \over 2}{1 \over d}{K_\theta}{\left( {\theta  - {\theta _0}} \right)^2}H\left( {{\rm{\Phi }} - {{\rm{\Phi }}_e}} \right),
\label{eq3}
\end{equation}

where $H(\Phi-\Phi_e)$ is the Heaviside step function to track scales engagement. The right-hand side of this equation can be considered as the summation of the energy absorbed by the substrate's elastic torsion $U_{substrate}={1 \over 2}{C_f}{C_w}{G_B}I{{\rm{\Phi }}^2}$, and the scales engagement $U_{scales}={1 \over 2}{1 \over {d}}{K_\theta}{( {\theta  - {\theta _0}})^2}H( {{{\Phi }} - {{{\Phi }}_e}})$. The torque-twisting rate relationship for the system could be found by differentiating Eq.~\ref{eq3} with respect to $\Phi$ and is written as:

\begin{equation}
T\left( {\rm{\Phi }} \right) = {C_f}{C_w}{G_B}I{\rm{\Phi }} + {{{K_\theta}} \over d}\left( {\theta  - {\theta _0}} \right){{\partial \theta } \over {\partial {\rm{\Phi }}}}H\left( {{\rm{\Phi }} - {{\rm{\Phi }}_e}} \right).
\label{eq4}
\end{equation}

\begin{figure}[htbp]
\centering
\subfloat[]{%
\includegraphics[scale = 0.34]{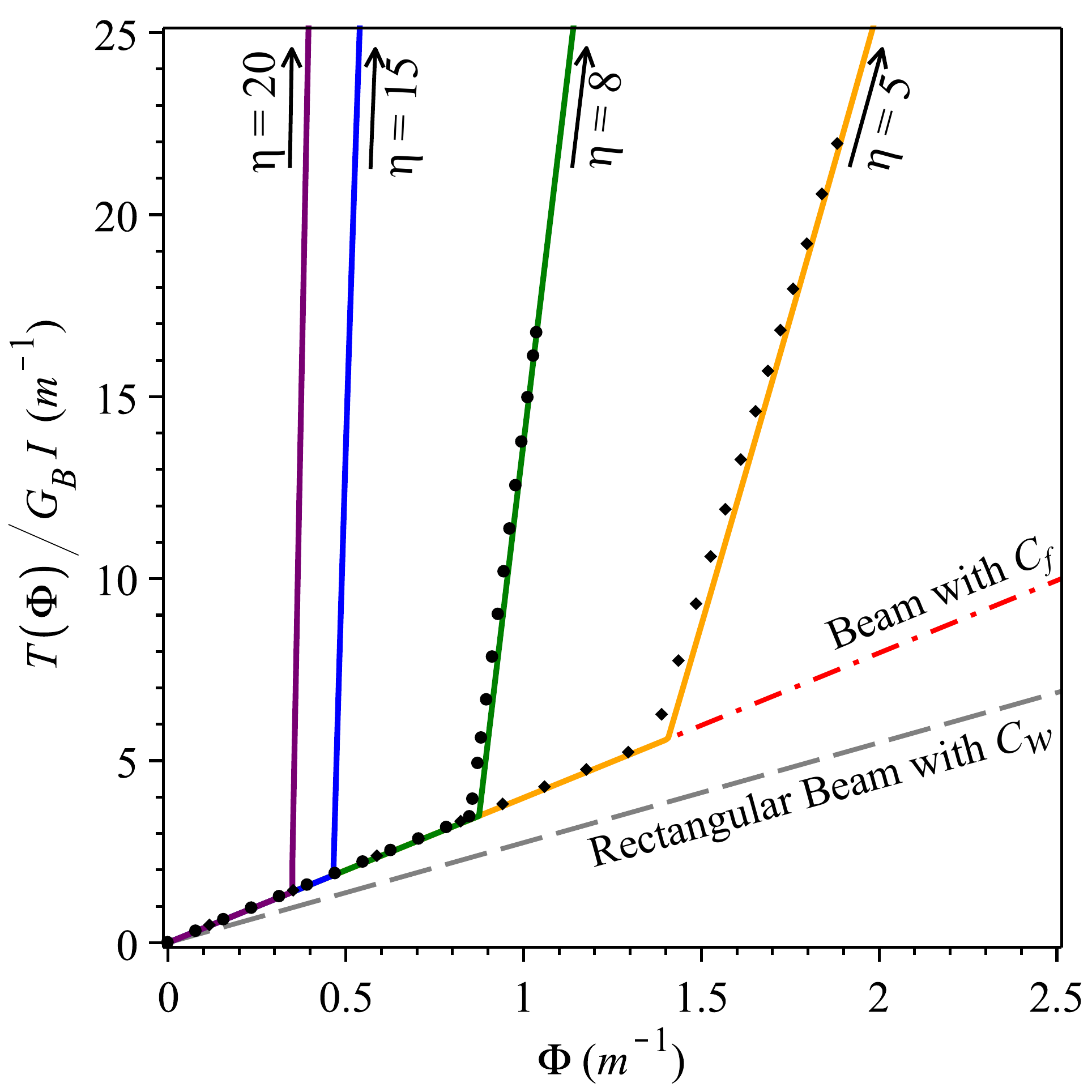}
\label{fig5A}}
\quad
\subfloat[]{%
\includegraphics[scale=0.3]{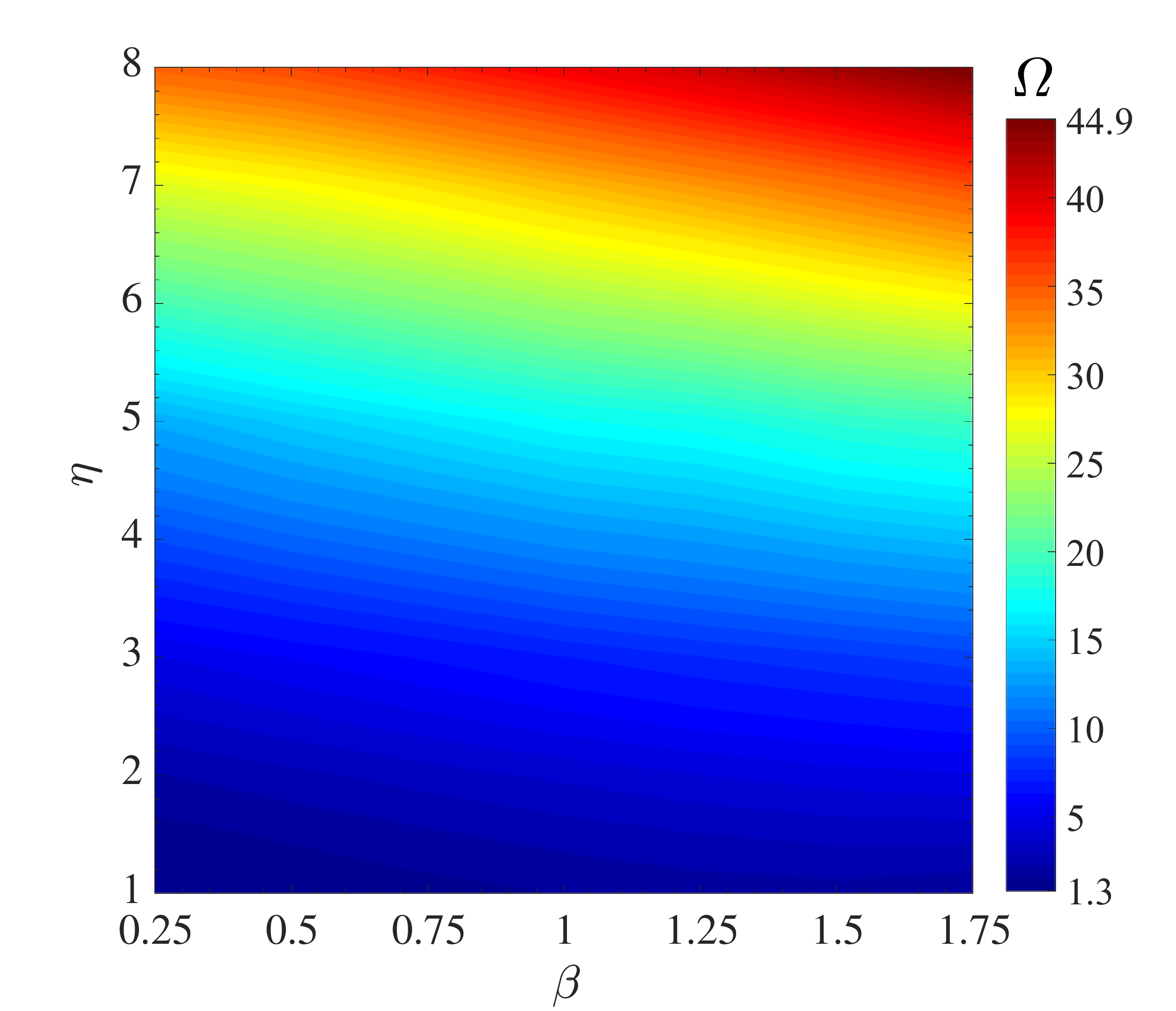}
\label{fig5B}}
\caption{(a) Phase map of dimensionless torque (${{T\left( {\rm{\Phi }} \right)} \over {{G_B}I}}$) versus twist rate ($\Phi$) for defined biomimetic structure for different $\eta$ with the given values of $\alpha=45^\circ$, $\theta_0=10^\circ$, $\beta=1.25$, and $\lambda=0.45$. Solid lines show analytical results and dotted lines are the representative of FE results. (b) Phase plot of the ratio between the total work done by the system, includes elastic torsion of the beam and scale engagement, and the elastic torsion work ($\Omega$), spanned by $\eta$ and $\beta$ with the given values of $\alpha=45^\circ$, $\theta_0=10^\circ$ and $\lambda=0.45$.}
\label{fig5} 
\end{figure}

This nonlinear expression is plotted in Fig.~\ref{fig5A} for different $\eta$ with $\theta_0=10^\circ$, $\alpha=45^\circ$, $\beta=0.6$, and $\lambda=0.32$. The properties of the substrate are considered as $G_B=10$ $GPa$ and $\nu=0.25$ with the cross section’s dimension of $32 \times 16$ $mm$. The scale spacing, the thickness and the embedded length of the scales are assumed as $d=25$ $mm$, $D=0.1$ $mm$ and $L=4.5$ $mm$. The results are compared to analogous FE simulations and we find a remarkably good fit with our model. The plot clearly demonstrates the sharp rise in nonlinear stiffening. The plot also highlights the inclusion effect in significantly increasing the torsional stiffness even before the engagement and underscores the accuracy of our model. We quantify the geometric tailorability of elastic energy of this system by using a magnification factor $\Omega=({U_{substrate}+U_{scales}}) / {U_{substrate}}$, which is the ratio of maximum possible energy absorbed by a substrate (from initial to lock) to an equivalent plain substrate. This factor is plotted in a phase map spanned by $\eta$ and $\beta$, Fig.~\ref{fig5B}, and shows the rapid increase of energy brought about by $\eta$ for any given $\beta$. However for a given $\eta$ the increase is relatively mild but still monotonic positive (see Table SI in ~\cite{c52}). This establishes the role of these parameters in both boosting stiffness and increasing energy, even though increasing $\eta$ leads to lower envelope of operation.

The linearized torque-twist for small $\theta$ is:  

\begin{equation}
{{T\left( {\rm{\Phi }} \right)} \over {{G_B}I}} = {C_f}{C_w}{\rm{\Phi }} + 2{C_B}\beta \left( {1 + \nu } \right){{{D^2}{d^2}} \over I}{\left( {{L \over D}} \right)^n}{\left( {\eta \tan \alpha  + \beta  - \sin \alpha } \right)^2}\left( {{\rm{\Phi }} - {{\rm{\Phi }}_e}} \right)H\left( {{\rm{\Phi }} - {{\rm{\Phi }}_e}} \right),
\label{eq5}
\end{equation}

where we recall that $G_B={E_B \over 2(1+\nu)}$. This linearized analytical expression sheds important light on the role of geometric parameters in enhancing the torsional stiffness of the substrate in small deformation. Particularly apparent is the effect of the lateral $\beta$ parameter, which has a nearly cubic relationship to torque. Therefore, increasing the width of the scales quickly increases rotational stiffness, significantly decreasing compliance in twisting. The effect of overlap ratio $\eta$ is quadratic, similar to bending behavior. This highlights the distinctness of the twisting response of the biomimetic scale-covered substrate. This plot also conforms with the more gentle slope of the experimental sample which corresponds to are $b=18$ $mm$, $l=26$ $mm$, $t=6.25$ $mm$, $d=14$ $mm$, $\alpha=45^\circ$ and $\theta_0=10^\circ$, which leads to $\eta=1.86$, $\eta=1.29$, $\lambda=0.45$. (see Fig. S5 in ~\cite{c52})

In conclusion, our work shows the geometrical tailorability of elastic response under twisting loads including stiffness, envelopes of operations and the overall energy landscape. We find that stiffness increase brought about by scales is highly nonlinear, reversible, rapid and tailorable, distinct from simply coating or embedding with a stiffer material or making a composite. This system exhibits a very specific nonlinear behavior which includes a seamless straddling between linear elastic, nonlinear elastic and finally a quasi-rigid behavior which exhibited by neither the PLA nor the silicone material. Each one of these regimes can be tailored using a different geometrical arrangement. This study strengthens the arguments of using biomimetic scales for designing structural metamaterials in a wide range of applications beyond simple bending. The current analytical model is aimed primarily to obviate the need for detailed fully resolved FE simulations for some aspects of design and analysis. These FE simulations become prohibitive for larger number of scale contacts, larger twists or for future work on dynamics, which would require repeated FE simulations on the structure.


\bibliography{MainFile}

\end{document}